\documentstyle[12pt,psfig]{article}  
\newcommand{\labell}[1]{\label{#1}}  
\newcommand{\reef}[1]{(\ref{#1})}

\def\IR{{\hbox{{\rm I}\kern-.2em\hbox{\rm R}}}} 
\def\IB{{\hbox{{\rm I}\kern-.2em\hbox{\rm B}}}} 
\def\IN{{\hbox{{\rm I}\kern-.2em\hbox{\rm N}}}} 
\def\IC{\,\,{\hbox{{\rm I}\kern-.59em\hbox{\bf C}}}} 
\def\IZ{{\hbox{{\rm Z}\kern-.4em\hbox{\rm Z}}}} 
\def\IP{{\hbox{{\rm I}\kern-.2em\hbox{\rm P}}}} 
\def\IH{{\hbox{{\rm I}\kern-.4em\hbox{\rm H}}}} 
\def\ID{{\hbox{{\rm I}\kern-.2em\hbox{\rm D}}}} 
\def\II{{\hbox{\rm I}\kern-.2em\hbox{\rm I}}}

\begin{document}  
  
\newpage  
\bigskip  
\hskip 4.7in\vbox{\baselineskip12pt  
\hbox{hep-th/0008081}}  
  
  
\bigskip  
\bigskip  
  
\centerline{{\Large \bf The Enhan\c con and ${\cal N}=2$ Gauge 
    Theory/Gravity RG Flows}} 
  
\bigskip  
\bigskip  
\bigskip  
  
\centerline{\bf Nick Evans{$^\flat$},  
 Clifford V. Johnson$^\natural$, Michela Petrini$^\sharp$}

\bigskip  
\bigskip  
\bigskip  
  
\centerline{\it $^\flat$Department of Physics,  
University of Southampton}  
\centerline{\it Southampton SO17 1BJ, U.K.}  
\centerline{$\phantom{and}$}  
\centerline{\it $^\natural$Centre  
for Particle Theory, Department of Mathematical Sciences}  
\centerline{\it University of  
Durham, Durham, DH1 3LE, U.K.}  
\centerline{$\phantom{and}$}  
\centerline{\it$^\sharp$Theoretical Physics Group, Blackett Laboratories}  
\centerline{\it Imperial College, London SW7 2BZ, U.K.}  
  
\centerline{$\phantom{and}$}  
  
\centerline{\small \tt n.evans@hep.phys.soton.ac.uk,  
  c.v.johnson@durham.ac.uk, m.petrini@ic.ac.uk}  
  
\bigskip  
\bigskip  
  
  
\begin{abstract}  
  \vskip 2pt We study the family of ten dimensional type~IIB
  supergravity solutions corresponding to renormalisation group flows
  from ${\cal N}=4$ to ${\cal N}=2$ supersymmetric $SU(N)$ Yang--Mills
  theory.  Part of the solution set corresponds to a submanifold of
  the Coulomb branch of the gauge theory, and we use a D3--brane probe
  to uncover details of this physics. At generic places where
  supergravity is singular, the smooth physics of the probe yields the
  correct one--loop form of the effective low energy gauge coupling.
  The probe becomes tensionless on a ring at finite radius.
  Supergravity flows which end on this ``enhan\c con'' ring correspond
  to the vacua where extra massless degrees of freedom appear in the
  gauge theory, and the gauge coupling diverges there.  We identify an
  $SL(2,\IZ)$ duality action on the enhan\c con ring which relates the
  special vacua, and comment on the massless dyons within them. We
  propose that the supergravity solution inside the enhan\c con ring
  should be excised, since the probe's tension is unphysical there.
\end{abstract}  
\newpage  
\baselineskip=18pt  
\setcounter{footnote}{0}  
  
  
\section{Introduction}  
There has been some emphasis on the construction of supergravity  
solutions which have a dual interpretation as  the  
renormalisation group flow from ${\cal N}=4$ supersymmetric pure  
Yang--Mills theory in the ultraviolet (UV) to ${\cal N}=2,1$ or 0  
supersymmetric Yang--Mills theories of various sorts, in the infrared  
(IR) \cite{gppz1}--\cite{pw2}.  
  
The supergravity solutions interpolate between AdS$_5\times S^5$, the 
dual of the ${\cal N}=4$ theory \cite{malda,gkp,w1}, which is at 
$r=\infty$ (where $r$ is a suitably chosen radial coordinate of 
AdS$_5$), and some much more complex solution in the interior, towards 
smaller $r$. Such solutions are typically  found in 5 dimensional 
supergravity, and in some cases they can be lifted to 10 dimensional 
type~IIB solutions. They often possess naked curvature singularities 
at finite values of $r$~\cite{gubby}. It is clear that some of these 
singularities are simply not physical, and the whole supergravity 
solution represents a flow to field theory with pathological 
behaviour, which presumably would be represented in terms of a 
pathology of the full type~IIB string theory solution. For other 
singularities, it is to be expected that the full stringy physics will 
resolve the singular physics into benign phenomena consistent with the 
dual gauge theory. 
  
The full technology for studying string theory in such backgrounds  
---where there is high curvature and strong Ramond--Ramond fields---  
is yet to be developed to a point where we can answer all of the  
questions raised by these issues.  However, while we wait for these  
techniques, we may try to use the tools already at our disposal for  
clues as to how the physics of such singularities is to be resolved.

In a context slightly different from the AdS/CFT correspondence,  
supergravity solutions  containing the physics of (but not fully  
dual to or decoupled from) pure ${\cal N}=2$ supersymmetric Yang--Mills  
theory were found to have naked, unphysical singularities \cite{jpp}.  
The singular solutions were not obtained as RG flows from some smooth  
supergravity dual of a more symmetric theory, but were deduced from  
properties of the gauge theory it was desired to represent, combined  
with knowledge of the parent string theory and the world--volume curvature  
couplings of the branes involved.

Using various clues and techniques, a method was found by which string 
theory excises the singular part of the solution in a way that was 
consistent with the gauge theory physics. This is called the ``enhan\c 
con mechanism'', and simply put, the idea is that the constituent 
branes making up the supergravity/string solution have expanded to 
form a spherical shell (called the ``enhan\c con'') around the region 
which appears singular in supergravity. The interior of this shell 
contains no point--like brane sources at all, and is (to a first 
approximation) a smooth, flat region of spacetime \cite{jpp}. 
  
A key tool in arriving at this conclusion was the idea of using a  
single  D--brane as a probe of the geometry created by the  
others. D--branes as probes of the ``true'' nature of spacetime at  
short distances and/or at strong string coupling, have been shown  
to be extremely useful, and when combined  
with ideas from gauge theory \cite{douglas} (which lives on the  
brane's world--volume \cite{joedidit,edbound}),  form part of a kit of  
sharp tools for studying the issues at hand \cite{shorty}.  
\footnote{For a collection of pedagogical studies with a focus on  
  these techniques, the reader may wish to have ref.\cite{primer} to  
  hand.}  
  
In this short note, we would like to turn to the issue of the 
supergravity ``duals'' of ${\cal N}=2$ supersymmetric Yang--Mills 
theory again, but now in the context of solutions describing RG flow 
from the ${\cal N}=4$ theory in the presence of appropriate masses. 
Recently, the full ten dimensional ``lift'' of the five dimensional 
supergravity solutions representing the flows \cite{gub2,pw1,bs2} has 
been presented \cite{pw1}. The point is that, just like in the case of 
ref.\cite{jpp}, the dual physics is not just that of supergravity, but 
of the full type~IIB string theory, and the idea here is to try to uncover 
some of the essential details of the crossover between the two. 
  
The ten dimensional solutions are generically singular in the IR, 
having naked curvature singularities, while being smooth in the UV, 
where they are weakly curved, matching onto AdS$_5\times S^5$, which 
is dual to the ${\cal N}=4$ supersymmetric $SU(N)$ Yang--Mills theory 
at large $N$.  We introduce a D3--brane probe and study the nature of 
the effective Lagrangian for moving the D3--brane probe in a 
supersymmetric way in the subspace corresponding to (a part of) the 
Coulomb branch of the ${\cal N}=2$ supersymmetric four dimensional 
gauge theory.  The result has the complementary interpretation as the 
effective geometry seen by the D3--brane probe as it moves in ten 
dimensional spacetime, and as the physics of the moduli space of the 
gauge theory on the constituent branes. 
  
We find that there {\it is} an enhan\c con shell, just as in the 
prototype cases studied in ref.\cite{jpp}, which in this case is a 
ring. Here, there is no enhanced gauge symmetry in the parent 
supergravity, (which was partly responsible for the name ``enhan\c 
con''), but all of the other key phenomena are present. In summary: 
  
\begin{itemize}  
  
\item There is a ring where the tension of the probe brane drops to  
zero, signalling the last radius where there is any meaning to the  
constituent branes as localised sources. The branes have spread out  
into a ring there and the supergravity solution interior to that  
---which is singular--- should be drastically modified by  
smoothing.

\item The full result for the spacetime metric as seen by the probe  
  interpolates smoothly between the simple behaviour at infinity and  
  that of the interior. In gauge theory language this results  
  in a  description of the behaviour of the gauge coupling  
  from the UV (where it is constant) to the IR (where it runs  
  logarithmically).\footnote{Since the work presented in  
    ref.\cite{jpp} was a study of the pure ${\cal N}=2$ gauge theory  
    in isolation, {\it i.e.} not connected to another gauge theory by  
    a massive continuous RG flow, the smooth interpolation to  the  
    physics of  the ${\cal N}=4$ gauge theory  
    is  not present there.}

\item Deep in the IR, the coupling of the gauge theory is seen to run  
  logarithmically, and it diverges at the enhan\c con ring,  
  representing the locus of points on the Coulomb branch where new  
  massless degrees of freedom appear in the gauge theory.  
  
\item The enhan\c con ring itself characterises a family of vacua at 
  which all species of $(p,q)$ dyons can become massless. There is a 
  descendant of the type~IIB string theory's $SL(2,\IZ)$ duality group 
  which acts naturally on and within the vacua, and the existence of 
  the dyons with this action follows from the existence of bound 
  states of type~IIB strings. There is a beautiful pattern of 
  relationships between the vacua as one goes around the circle. 
  
\end{itemize}

It is very encouraging to see the enhan\c con behaviour appearing so 
naturally in the supergravity flow arena, and we expect that this 
technique will find further fruitful application in this context as 
more ten dimensional supergravity solutions are found. 
  
\section{The Ten Dimensional Solution}  
\label{tendee}  
The family of ten dimensional solutions of ref.\cite{pw1} describing 
the gravity dual of ${\cal N}=4$ supersymmetric $SU(N)$ Yang--Mills 
theory, broken to ${\cal N}=2$ in the IR may be written as: 
\begin{equation}  
ds^2_{10}=\Omega^2 ds^2_{1,4}+ds^2_{5}\ ,  
\labell{fullmetric}  
\end{equation}  
for the Einstein metric, where  
\begin{equation}  
ds^2_{1,4}=e^{2A(r)}\left(-dt^2+dx_1^2+dx_2^2+dx_3^2\right)+dr^2\ ,  
\labell{littlemetric}  
\end{equation}  
and  
\begin{equation}  
ds_5^2={L^2}{\Omega^2\over\rho^2}\left[{d\theta^2\over  
    c}+\rho^6\cos^2\theta\left({\sigma_1^2\over c  
    X_2}+{\sigma_2^2+\sigma_3^2\over X_1}\right)  
+\sin^2\theta {d\phi^2\over X_2}\right]\ ,  
\labell{bigmetric}  
\end{equation}  
with $c=\cosh (2\chi)$, and  
\begin{eqnarray}  
\Omega^2&=&{(c X_1 X_2)^{1/4}\over \rho}\nonumber\\  
X_1&=&\cos^2\theta+\rho^6\cosh(2\chi)\sin^2\theta\nonumber\\  
X_2&=&\cosh(2\chi)\cos^2\theta+\rho^6\sin^2\theta\ .  
\labell{warp}  
\end{eqnarray}  
The $\sigma_i$ are the standard $SU(2)$ left--invariant forms, the sum  
of the squares of which would give the standard metric on a round  
three--sphere.

The functions $\rho(r)=e^{\alpha(r)}$ and $\chi(r)$ which appear in 
the lifted 10 dimensional metric are the 5 dimensional supergravity 
scalars pertaining to certain operators in the dual gauge theory. 
There is a one--parameter family of solutions for them, giving a family 
of supergravity solutions, and correspondingly a slice through a {\it 
  moduli space} of ${\cal N}=2$ theories in the IR. We will discuss 
$\rho(r)$ and $\chi(r)$ further shortly. 
  
At $r\to\infty$, the various functions in the solution have the 
following asymptotic behaviour: 
\begin{equation}  
\rho(r)\to 1\ ,\,\, \chi(r)\to 0\ ,\,\, A(r)\to r/L\ ,  
\labell{asymptotes}  
\end{equation}  
where $L=\alpha^{\prime1/2} (2g^2_{\rm YM}N)^{1/4}$.  Also $g^2_{\rm 
  YM}=2\pi g_s$, here. 
  
It is easily seen that the non--trivial radial dependences of 
$\rho(r)$ and $\chi(r)$ deform the supergravity solution from 
AdS$_5\times S^5$ at $r=\infty$ where there is an obvious $SO(6)$ 
symmetry (the round $S^5$ is restored), to a spacetime which only has 
an $SU(2)\times U(1)^2$ symmetry, which is manifest in the metric 
\reef{bigmetric}. 
  
There are also explicit solutions given in ref.\cite{pw1} for the 
dilaton, $\Phi$, the R--R scalar and two--form potential, $C_{(0)}$ and 
$C_{(2)}$, and the NS--NS two--form potential $B_{(2)}$.  In fact, for 
the subspace of the solutions which we will probe in this paper (see 
later), the two--form fields are zero, and so we shall not list their 
full form here. 
 
The fields $(\Phi, C_{(0)})$ are gathered into a complex scalar field 
which we shall denote as $\lambda=C_{(0)}+ie^{-\Phi}$. This is a 
natural object in the basis where the classical $SL(2,\IR)$ symmetry 
of the type~IIB supergravity is manifest. However, in ref.\cite{pw1} the 
more venerable $SU(1,1)$ basis is used, and the supergravity field $B$ 
given there is related to $\lambda$ as follows: 
\begin{equation} 
\lambda=i\left({1-B\over 1+B}\right)\ . 
\labell{axidill} 
\end{equation} 
The explicit solution is
\begin{equation} 
B=\left[{b^{1/4}-b^{-{1/4}}\over b^{1/}+b^{-1/4}}\right]e^{2i\phi}\
,\,\,\,
{\rm 
  where}\,\,\, 
b\equiv c{X_1\over X_2}\ . 
\labell{relations} 
\end{equation}

In fact, we shall not yet need the explicit form for the dilaton 
because it disappears from the D3--brane probe action when it is 
written in Einstein frame, as we shall do in section \ref{probing}. 
The scalar field $C_{(0)}$ couples to $F\wedge F$ on the D3--brane 
world volume, contributing to the $\theta$--angle in the ${\cal N}=2$ 
effective low energy theory. 

We will need the explicit form for the R--R four form potential  
$C_{(4)}$, to which the D3--brane naturally couples. It  
is:~\footnote{We noticed that the expression given in equation (4.9)  
  of ref.\cite{pw1} is not consistent with their equation (3.31), and  
  we assume that it is a typographical error. There is a crucial  
  factor of $\sinh^2(2\chi)$ missing. Instead of checking the ten  
  dimensional solution by hand, we verify this correction by requiring  
  that the potential seen by the brane probe is constant, as required  
  by  supersymmetry. We have also inserted a factor of  
  $g_s^{-1}$ to match our standard conventions for the D3--brane  
  charge.}  
\begin{equation}  
C_{(4)}=  
e^{4A}{X_1\over g_s\rho^2} dx_0\wedge dx_1\wedge dx_2 \wedge dx_3\ .  
\end{equation}  
  
The radial dependences of the various functions which appear in the  
solution are given by the $5d$ supergravity equations of motion, written in  
the following form (recall that $\rho=\exp(\alpha)$):  
\begin{eqnarray}  
{d\alpha\over dr}&=&{g\over 6}\left({1\over\rho^2}  
-\rho^4\cosh(2\chi)\right)\nonumber \\  
{d\chi\over dr}&=&-{g\over 4}\rho^4\sinh(2\chi)\nonumber \\  
{dA\over dr}&=&{g\over 3}\left({1\over\rho^2}  
+{1\over2}\rho^4\cosh(2\chi)\right)\ ,  
\labell{diffys}  
\end{eqnarray}  
(where $g=2/L$) with the partial result\cite{pw1,bs2}:  
\begin{eqnarray}  
e^{A}&=&k{\rho^2\over\sinh(2\chi)}\nonumber\\  
\rho^6&=&\cosh(2\chi)+\sinh^2(2\chi)\left(\gamma  
+\log\left[{\sinh\chi\over\cosh\chi}\right]\right)\ .  
\label{partialresult}  
\end{eqnarray}  
Here, $k$ is a constant we shall fix later.   
 
More importantly, $\gamma$ is a constant whose values characterise a 
family of different solutions for $(\rho(r),\chi(r))$ representing 
different flows to the ${\cal N}=2$ gauge theory in the IR. (See figure 
\ref{gammas}, on page \pageref{gammas}.) 
 
To date, it has been hard to extract the physics of the ${\cal N}=2$ 
gauge theory from this solution. The authors of \cite{gub2, pw1} have 
proposed that the solutions with $\gamma \leq 0$ describe the gauge 
theory at different points on moduli space with the $\gamma = 0$ curve 
describing the singular point on moduli space where the gauge coupling 
diverges.  The functional dependence of the gauge coupling on moduli 
space has not been reproduced so far, and we present it in this paper. 
For these solutions, the five dimensional supergravity potential is 
bounded above by the asymptotic UV value, and this is suggested in 
ref.~\cite{gub2} as a phenomenological criterion for physical 
acceptability.

Strong evidence for the above interpretation of the $\gamma=0$ flow 
was obtained in ref.\cite{ep}. There, the five dimensional criterion 
of ref.\cite{gub2}, applied to solutions in the theory perturbed to 
${\cal N}=1$ gauge theory shows that the $\gamma=0$ solution is 
distinguished, as it yields the vacuum which is {\it not} lifted after 
the ${\cal N}=1$ perturbation.  
  
It would be nice to have a fully ten--dimensional criterion for what 
constitutes a physical flow, and our probe analysis of this paper is a 
concrete step in this direction.  In the calculations we present 
below, the physical function describing the running of the gauge 
coupling will emerge naturally from the solutions, and we will confirm 
the above identification of these solutions.

\section{The Gauge Theory}  
 \label{gaugetheory} 
The ${\cal N}=4$ supersymmetric Yang--Mills theory's gauge multiplet  
has bosonic fields $(A_\mu, X_i)$, $i=1,\ldots,6$, where the  
scalars $X_i$ transform as a vector of the $SO(6)$ R--symmetry, and  
fermions $\lambda_i$, $i=1,\ldots,4$ which transform as the $\bf 4$ of  
the $SU(4)$ covering group of $SO(6)$.  
  
In ${\cal N}=1$ language, there is a vector supermultiplet 
$(A_\mu,\lambda_4)$, and three chiral multiplets made of a fermion and 
a complex scalar ($k=1,2,3$): $$\Phi_k\equiv(\lambda_k, 
\phi_k=X_{2k-1}+iX_{2k})\ .$$ 
In this language, the  ${\cal N}=2$ 
supersymmetric Yang--Mills theory has the vector multiplet and one 
massless chiral multiplet, which we can choose to be $\Phi_3$.  The 
flow from the ${\cal N}=4$ theory to the ${\cal N}=2$ theory is 
therefore achieved by turning on operators which correspond to giving 
a mass to the other multiplets. As one falls well below the scale of 
these masses ---by going to the IR--- the theory becomes more 
effectively the theory we seek (although one must always remember that 
the gauge coupling is strongly coupled in the far UV in the cases 
where the supergravity is valid). 
  
The operators are switched on in supergravity by considering  a  
combination of the operators\cite{gub2,pw1,bs2}:  
\begin{eqnarray}  
\alpha:\qquad&& \sum_{i=1}^4{\rm Tr} (X_iX_i)-2\sum_{i=5}^6{\rm Tr}  
(X_iX_i)  
\nonumber\\  
\chi:\qquad&& {\rm Tr} (\lambda_1\lambda_2+\lambda_2\lambda_2)+{\rm h.c.}  
\end{eqnarray}  
The dictionary of the AdS/CFT correspondence assigns two specific 
scalars $(\rho=\exp(\alpha)$, $\chi)$ to these operators. The IR/UV 
property \cite{iruv} of the AdS/CFT correspondence then requires a 
non--trivial solution for these fields, making them functions of $r$ 
varying along the flow from the UV ($r=+\infty$) to the IR 
$r=-\infty$. Consistency of the supergravity equations of motion 
require that there be a non--trivial back--reaction on the geometry, 
giving a deformation of the spacetime metric, represented by $A(r)$, 
{\it etc.,} in section \ref{tendee}.

Moving around on the accessible part of the Coulomb branch of the 
${\cal N}{=}2$ theory corresponds to giving a vacuum expectation value 
(vev) to $\phi_3=X_5+iX_6$.  We shall explore this branch by moving a 
probe brane in those directions, and examining its physics. In the 
supergravity solution written  in section \ref{tendee}, this 
corresponds to moving in the $(r,\phi)$ plane, setting $\theta=\pi/2$. 
Attempting to move the probe out of this plane induces a non--trivial 
potential for the brane's motion indicating that such motion is not a 
modulus of the field theory. In fact, in the field theory it would 
correspond to inducing vevs for the massive scalars which is 
energetically disfavoured.

The Coulomb branch of the moduli space of the ${\cal N}=2$ $SU(N)$ 
gauge theory is parameterised by the vevs of the complex adjoint 
scalar $\phi_3$ which set the potential ${\rm 
  Tr}[\phi_3,\phi_3^\dagger]^2$ to zero. This generically breaks the 
theory to $U(1)^{N-1}$. This moduli space is of course an $N-1$ 
complex dimensional space. The low energy effective action of the 
theory is described in terms of the $N-1$ complex low energy fields 
$\phi_a$, with a matrix of couplings and $\theta$--angles given in 
terms of the complex couplings 
$\tau^{ab}(\phi_a)$~\cite{seib,seibwitt,af,klemm,dougshenk}. The 
theory is invariant under an $Sp(2N-2,\IZ)$ group of {\it duality} 
transformations which shifts and inverts the coupling matrix 
$\mathbf\tau$, and exchanges the fields $\phi_a$ with dual fields 
$\phi_a^D$.  This operation, which includes a strong--weak coupling 
duality, is crucial in the study of the physics of the Coulomb 
branch, as first demonstrated in ref.\cite{seibwitt}. 
 
We have only a one complex dimensional subspace of the moduli space 
here. As such, we have one complex field whose vev we wish to study 
and one complex coupling $\tau$ which depends on this vev.  We should 
expect that there is a subgroup of the $Sp(2N-2,\IZ)$ duality acting 
on our low energy theory, and we need not look far for its origin: Our 
moduli space is the physics of a single D3--brane probe, moving in the 
$\theta=\pi/2$ plane, and there is a $U(1)$ gauge theory living on it 
which shall be our effective low energy theory. There is a gauge 
coupling $g^2_{\rm YM}$ and a $\theta$--angle $\Theta$, which combine 
into a natural complex coupling 
\begin{equation} 
\tau={\Theta\over 2\pi}+i {4\pi\over g^2_{\rm YM}}\ . 
\end{equation} 
 
There is a natural $SL(2,\IZ)$ invariance of the world--volume action 
of the probe\cite{gibbons,green,tseytlin}, which is the self duality 
symmetry of the type~IIB superstring theory. This duality inverts and 
shifts the complex type~IIB coupling, which is the scalar $\lambda$ 
described in section \ref{tendee}, according to 
\begin{equation} 
\lambda\to {a\lambda+b\over c\lambda+d}\ ;\quad ad-bc=1\;\quad 
\{a,b,c,d\}\in \IZ\ . 
\end{equation}  It leaves the ten dimensional 
Einstein metric and the R--R four--form invariant, and mixes the
NS--NS and the R--R two--forms, under which the fundamental string and
D1--brane are charged. The resulting $(p,q)$ bound state
strings\cite{john,edbound} will play a role later. On the world--volume of
the brane, the duality exchanges the gauge field $F_{ab}$ with a dual
gauge field, $F^D_{ab}\equiv -2(\delta S/\delta F^{ab})$, where $S$ is
defined in equation~\reef{actiontime}.
 
This $SL(2,\IZ)$ action descends to an action on the effective low 
energy theory which we shall derive on the probe, with complex 
coupling $\tau$, which we will see the first hints of below. This 
duality transformation which acts on our moduli space, and it will be 
exciting to explore its uses further: It ought to give weakly coupled 
dual descriptions of the physics near the enhan\c con, where as we 
shall see, $g_{\rm YM}$ diverges.\footnote{The issue of the search for 
  a weakly coupled effective description of the enhan\c con is a 
  matter discussed in ref.\cite{me}.} A more thorough exploration of this 
descendant of $SL(2,\IZ)$ duality shall be left for future work. 
  
\section{Probing with a D3--brane}  
\label{probing}  
The uplifted geometry presented in ref.\cite{pw1} and listed in section  
\ref{tendee} is given in the Einstein frame.  It therefore makes sense  
to write the D3--brane world--volume action in terms of this:  
\begin{eqnarray}  
S=-\tau_3\int_{{\cal M}_4} d^4\xi\,\,  
{\rm det}^{1/2}[{G}_{ab}+e^{-\Phi/2}{\cal  
  F}_{ab}]+\mu_3\int_{{\cal M}_4} \left(C^{(4)}+{1\over2}C_{(0)}\, {\cal 
  F}\wedge {\cal F}\right), 
\label{actiontime} 
\end{eqnarray}  
where ${\cal F}_{ab}=B_{ab}+2\pi\alpha^\prime F_{ab}$, and ${\cal 
  M}_4$ is the world--volume of the D3--brane, with coordinates 
$\xi^0,\ldots,\xi^3$. The parameters $\mu_3$ and $\tau_3$ are the 
basic\cite{gojoe} R--R charge and tension of the D3--brane: $\mu_3 = 
\tau_3 g_s= (2\pi)^{-3}(\alpha^\prime)^{-2}$. Also, $G_{ab}$ and 
$B_{ab}$ are the pulls--back of the ten dimensional metric (in 
Einstein frame) and the NS--NS two--form potential, respectively. The 
String frame metric is ${\widetilde G}_{\mu\nu}=e^{\Phi/2}G_{\mu\nu}$, 
and the pull--back is {\it e.g.}: 
\begin{equation}  
G_{ab}=G_{\mu\nu}{\partial x^\mu \over\partial\xi^a}  
{\partial x^\nu \over\partial\xi^b}\ .  
\end{equation}  
  
We will work in static gauge and partition the spacetime coordinates, 
$x^{\mu}$, as follows: $x^i=\{x^0,x^1,x^2,x^3\}$, and 
$y^m=\{r,\theta,\phi,\varphi_1,\varphi_2,\varphi_3\}$. (The $\varphi_i$ 
are angles on the deformed $S^3$ of section \ref{tendee}.)  Static 
gauge is chosen as: 
\begin{equation}  
x^0\equiv t=\xi^0\ ,\quad x^i=\xi^i\ ,\quad  y^m=y^m(t)\ .  
\end{equation}

Focusing on the subspace $\theta=\pi/2$, for the ${\cal N}=2$ theory's  
Coulomb branch (as discussed in the previous section) we get the  
following result for the effective Lagrangian for the {\it slowly  
  moving} probe moving in the transverse directions $y^m=(r,\phi)$ (we  
restrict ourselves to considering $F_{ab}=0$ here):  
\begin{equation}  
{\cal L}={\tau_3\over2}\Omega^2 e^{2A}\sum_m G_{mm} {\dot y}^m{\dot y}^m\ .  
\end{equation}  
We have neglected terms higher than quadratic order in the velocities.  
There is no potential term; it vanished due to a cancellation between  
the leading term in the expansion of the determinant, and the term  
containing $C^{(4)}$. This result is consistent with the fact that we  
have eight supercharges. (See footnote 3.)  Writing this out  
explicitly, we see that:  
\begin{equation}  
{\cal L}={\mu_3\over g_s} \left({1\over 2}\Omega^4 e^{2A}  
{\dot r}^2+{L^2\over 2}  
{\Omega^4 e^{2A}\over \rho^8} {\dot\phi}^2\right)\ .  
\labell{theresult} 
\end{equation}  
  
Next, we should attempt to find good coordinates which are appropriate 
to the description of the physics of the Coulomb branch of an ${\cal 
  N}=2$ gauge theory. 
  
\section{The Coulomb Branch in the  UV limit}  
  
Now before proceeding further, we should check that our result makes  
sense in the UV limit ($r\to\infty$). There, we have $\Omega^4\to 1$,  
from \reef{asymptotes} and defining  
\begin{equation}  
 r=L \log \left(u{\alpha^{\prime}\over L}\right)\ , \,\, {\rm for}\,\,\, L= 
 \alpha^{\prime1/2} (2g^2_{\rm YM}N)^{1/4}\ , 
\end{equation}  
where $u$ has dimensions of energy, we have:  
\begin{equation}  
{\cal L}= {1\over 8\pi^2 g^2_{\rm YM}}  
\left( {\dot u}^2+u^2{\dot\phi}^2\right)\ .  
\labell{flat} 
\end{equation}  
  
This is the right result. We are looking at the result for a single 
D3--brane probing a two dimensional subspace of the full six 
dimensional transverse direction to $N$ other branes. The full result 
is flat, as it should be for the ${\cal N}=4$ theory. Interpreted in 
terms of gauge theory, $u$ is an energy scale, or the modulus of the 
vev of $\phi_3$. We see that there is no running for $g_{\rm 
  YM}$; there is a flat metric on moduli space. 
  
We can write our result in the expected  form\cite{seib,seibwitt}: 
\begin{equation} 
ds^2\sim {}{\rm Im}\tau \,\, dU d{\bar U}\ , 
\labell{expected} 
\end{equation} 
showing that the complex plane $(u,\phi)$ constitutes a good choice of 
coordinates for the ${\cal N}=2$ Coulomb branch in the UV. We shall 
have to work harder to find such good coordinates in the IR,  as 
is clear from the general expression in equation \reef{theresult}.

\section{The Coulomb Branch in the IR limit}  
  
Let us use the same radial variable $u$ as we did before, initially, 
although we will define another shortly. The idea should be that we 
simply see a deviation from the flat metric \reef{flat} as we come 
into smaller $u$. 
  
Since in the case $\theta=\pi/2$ we have:  
\begin{equation}  
\Omega^4=\rho^4\cosh 2\chi\ ;\quad e^{2A}={k^2\rho^4\over\sinh^2  
  (2\chi)}\ ,  
\labell{simplify} 
\end{equation}  
we can simply write the moduli space metric  as  
\begin{eqnarray}  
ds^2&=&   
{ L^2\Omega^4 e^{2A}\over 8\pi^2 g^2_{\rm YM}u^2\rho^8\alpha^{\prime2}}  
\left(\rho^8d{u}^2+u^2d{\phi}^2\right)\nonumber\\ 
&=& 
{1\over 8\pi^2 g^2_{\rm YM} }  
  { L^2\over\alpha^{\prime2} }{ k^2 \cosh (2\chi) \over u^2\sinh^2 (2\chi)}  
\left(\rho^8d{u}^2+u^2d{\phi}^2\right)  \ .  
\end{eqnarray}

Our next goal is to try to understand the physics of this metric for 
arbitrary distance into the IR. It is convenient\footnote{We thank Rob 
  Myers for a crucial comment.} to define a new radial parameter, 
$v=u/h$, such that $h(u)\to 1/a$ as $u\to\infty$, where $a$ is a 
dimensionless constant. 
Furthermore, if we choose $\rho^4(u)= h(u) dv/du$, our metric takes 
the simple form: 
\begin{equation}  
ds^2=  
{1\over 8\pi^2 g^2_{\rm YM} }  { L^2\over\alpha^{\prime2} } 
  { k^2 \cosh (2\chi) \over v^2\sinh^2 (2\chi)}  
\left(d{ v}^2+v^2d{\phi}^2\right)  \ . 
\labell{verynice}  
\end{equation}  
In this form we can interpret the function of $v$ outside the 
canonical Lagrangian as the running of the coupling ${\rm Im}\tau$ in 
the ${\cal N}=2$ theory.  The next steps are to study the behaviour of 
the metric component: 
\begin{equation} 
G_{vv}(v)\equiv{1\over 8\pi^2 g^2_{\rm YM}(v)} = 
{N\over 4\pi^2 }  
{ k^2 \cosh (2\chi(v))\over L^2v^2\sinh^2 (2\chi(v))}  
\labell{component} 
\end{equation} 
as a function of $v$, and also to find a suitable\cite{seib} local 
presentation of the effective action at least at low energy.  Note now 
that $\chi(v)$ satisfies a different differential equation. Since 
$$  
{d\chi\over dr}={d\chi \over du}{du\over dr}={d\chi \over dv}{dv\over  
  du}{u\over L}={d\chi\over dv}\rho^4{v\over L}\ ,  
$$ a new equation may be deduced from \reef{diffys}:  
\begin{equation}  
{d\chi\over dv}=-{1\over 2v}\sinh(2\chi)\ .  
\end{equation}  
remarkably, this is an easy equation to solve, and the result is:  
\begin{equation}  
\chi(v)={1\over 2}\log\left[  
\tanh\left({1\over 2}\log\left({v\over v_e}  
\right)\right)\right]\ . 
\labell{ooohlovely}  
\end{equation}  
where $v_e$ is an integration constant!

The truly remarkable thing is that it is easy to see that the function 
$G_{vv}(v)$ has a zero on the circle $v=v_e$, which deserves to be 
called the ``enhan\c con'', since\cite{jpp} the probe's tension is 
going to zero there! 
 
\begin{figure}[ht]  
\centerline{\psfig{figure=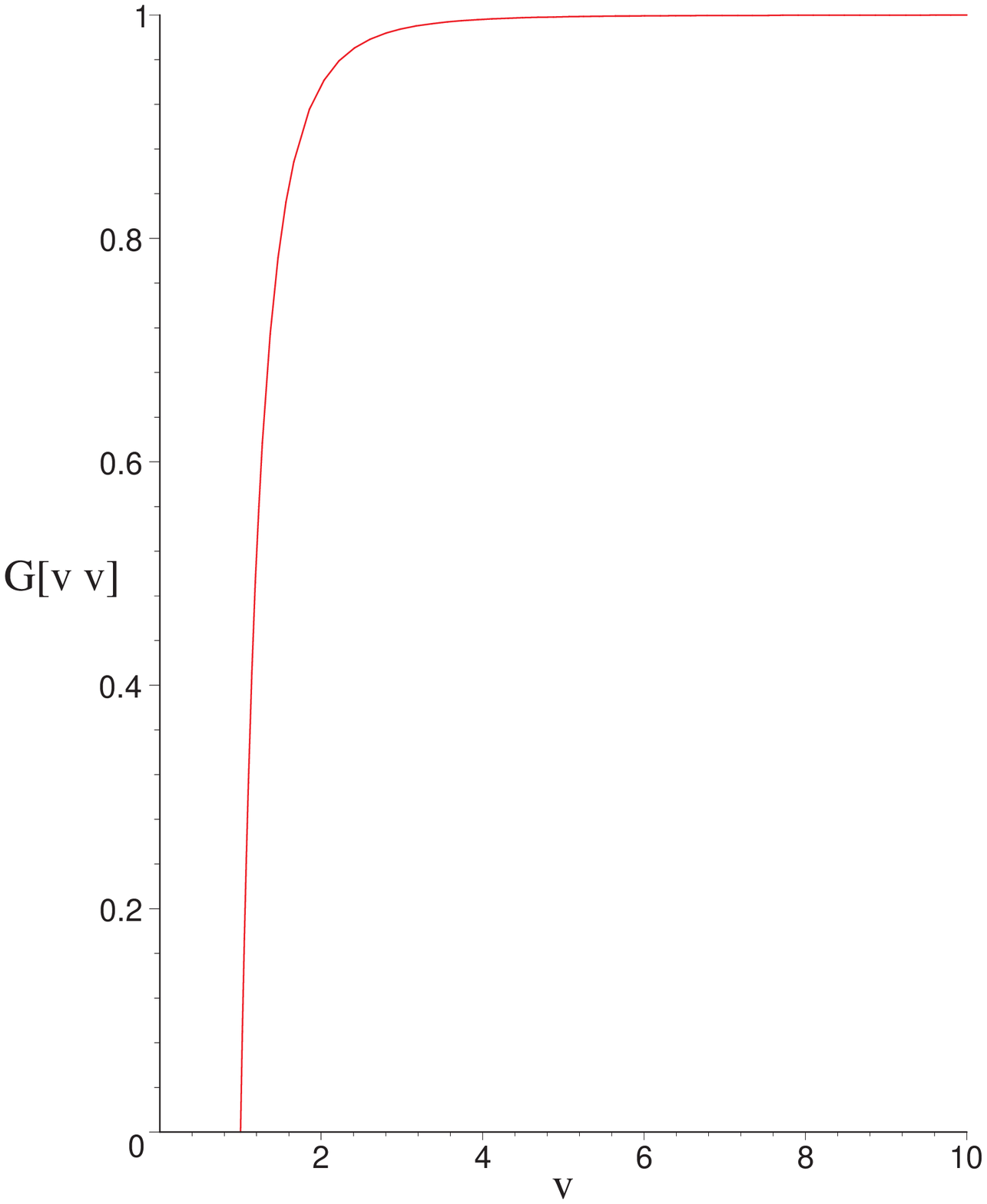,height=2.6in,width=2.7in} 
\psfig{figure=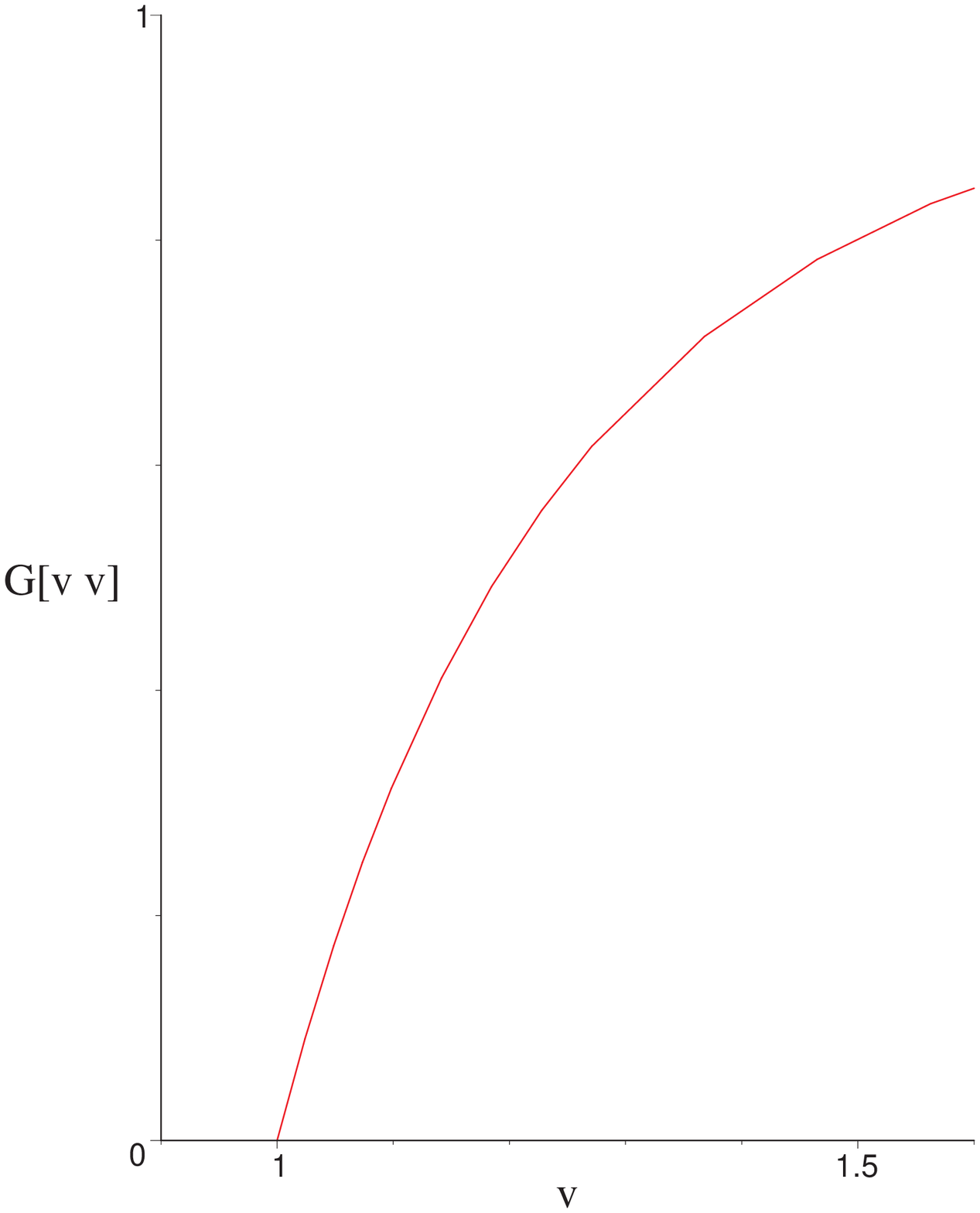,height=2.6in,width=2.7in}}  
\caption{\small The metric function $G_{vv}$ representing the moduli space  
  probe result for the D3--brane. It is proportional to the probe's 
  tension. The enhan\c con is at $v_e=1$ here. This is also the result 
  for the running of the inverse squared Yang--Mills gauge coupling as 
  a function of energy scale $v$. It stops running in the UV. On the 
  right is a closeup, showing the logarithmic running near $v_e$, in the IR.} 
\label{probemetric}  
\end{figure}

(Note also that $\chi(v)$ has a mild divergence at $v=v_e$, and is 
smooth and exists for $v>v_e$.)  The behaviour of $G_{vv}(v)$ is 
displayed in figure \ref{probemetric}.  (There, we have chosen various 
constants to be unity for plotting convenience.)  The first thing to 
notice is that our function correctly goes to unity in the UV limit, 
rapidly becoming nearly constant, showing the (near) scale invariance 
of the theory. It fact, this nearly conformal behaviour dominates much 
of the flow, which is interesting. 
More engaging behaviour can be seen in the neighbourhood of 
$v=v_e$, where we may expand the function to see that in that region 
(we fix $k=2v_e\alpha^\prime/(aL)$ by 
examining the UV asymptotics and also equation \reef{simplify}): 
\begin{equation}  
ds^2\sim
{N}\log\left({v\over v_e}\right)  
\left(d{v}^2+v^2d{\phi}^2\right)  \ .  
\labell{logarithm}  
\end{equation}  
which is the one--loop logarithmic running we expect from the gauge
coupling of a $U(1)$ low energy theory on the Coulomb branch of the
${\cal N}{=}2$ gauge theory! 
 
(It is important to note that we have set $a^{-2}=g^2_{\rm YM}N$ by 
hand. One would expect to be able to fix $a$ uniquely in terms of the 
supergravity parameters in the solution, and that we have not quite 
done this suggests that our attempt to adjust the conventions of 
ref.\cite{pw1} to restore all factors of $N$ and $g_{\rm YM}$ is 
incomplete. However, we are confident that our parameters are correct 
for the following reasons: On general grounds\cite{malda}, the 
complete decoupling from gauge theory of the dual supergravity ensures 
that it must depend on $g_{\rm YM}$ and $N$ {\it only} through the 
t'Hooft coupling $\lambda=g^2_{\rm YM}N$. So our choice for $a$ is 
not unnatural, up to pure numbers.  Furthermore, the structure of the 
AdS/CFT correspondence is such that we should get the one--loop field 
theory result in this way, which fixes the factor of $g_{\rm YM}$, and 
then the $N$ {\it must} accompany it to make $\lambda$.  See also 
ref.\cite{jpp}.)

Quite pleasingly, we have succeeded in showing analytically that our 
metric (and hence our low energy effective action) can be written in 
the form: 
\begin{equation} 
ds^2\sim {}{\rm Im}\tau(V) dV d{\bar V}\ , 
\end{equation} 
where now the complex coordinate $V$ is the $(v,\phi)$
plane.\footnote{{\it Added in v2 of preprint:} One can check that our
  function agrees with the coefficient of the $F^2$ term in the
  effective action on the probe.  That coefficient is simply
  $e^{-\Phi}$ and from \reef{ooohlovely} and \reef{axidilaton}, it can
  be seen that its IR limit is also proportional to $\log(v/v_e)$. The
  dependence of the coefficient on $\phi$ can be absorbed by a
  rescaling of $v$.  There is $\phi$ dependence in the coupling since
  the $U(1)$ symmetry on the ${\cal N}=2$ theory's moduli space is not
  an exact symmetry of the broken ${\cal N}=4$ theory, especially
  when, as in this case, the ${\cal N}=4$ matter is not decoupled at
  the strong coupling scale. See ref.\cite{bpp} for a change of
  coordinates which emphasises this, while converting the circular
  enhan\c con into a line segment by squashing it.}  The full metric
interpolates between this form of coordinates in the IR and the
similar form~\reef{expected} found earlier in terms of the the
coordinate $U$ in the UV.
 
We are now ready to discuss the interpretation and consequences of our 
computation for the Coulomb branch of the gauge theory.

\section{Families of Flows: The Moduli Space}  
 
\begin{figure}[ht]  
\centerline{\psfig{figure=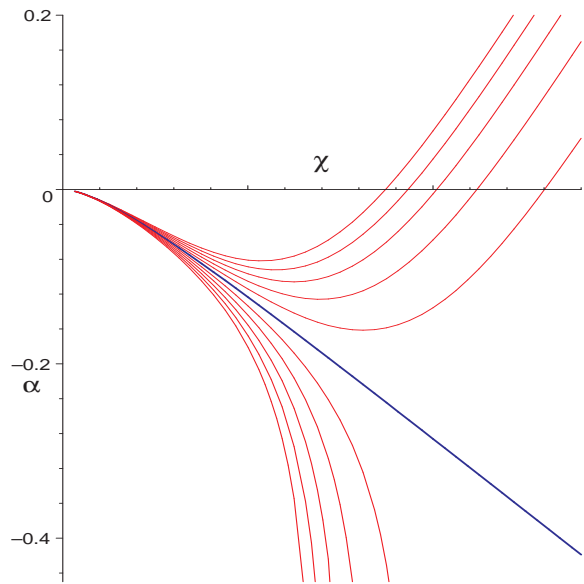,height=3.0in,width=3.0in}}  
\caption{\small The families of $(\chi,\alpha)$ curves for differing 
  $\gamma$, given by equation \reef{partialresult}. There are three 
  classes of curves. The blue (middle) curve is $\gamma=0$, the 
  $\gamma<0$ curves are below it, and the $\gamma>0$ curves are above. 
  The flow from UV to IR along each curve is to the right. Recall that 
  $\rho=e^\alpha$, and refer to the text for further physical 
  interpretation of each curve. } 
\label{gammas}  
\end{figure}

As mentioned before, the ${\cal N}=2$ flows are characterised into 
three different classes $\gamma<0$, $\gamma=0$ and $\gamma>0$ (see 
figure \ref{gammas}, on page \pageref{gammas}).  Equation 
\reef{ooohlovely} and the metric \reef{logarithm} must be interpreted 
carefully for each of these flows.

Let us first check the supergravity behaviour of these flows: 
\begin{itemize} 
\item For $\gamma<0$, equation \reef{partialresult} yields a {\it 
    finite} value of $\chi$ in the IR, while $\rho$ goes to {\it 
    zero}.  It is clear, from the functions listed at the beginning of 
  section \ref{tendee}, that the ten dimensional supergravity solution 
  has a naked singularity as a result.

\item For $\gamma=0$, $\chi$ {\it diverges} in the IR and again $\rho$ 
  goes to zero. Supergravity again has singular behaviour, coming from 
  both the divergence and the zero. 
   
\item For $\gamma>0$ both $\chi$ and $\rho$ diverge, and the 
  supergravity is singular. 
 
\end{itemize}

Let us now see how this behaviour combines with our
equations~\reef{ooohlovely} and~\reef{logarithm} to give sensible
physics.

\begin{itemize}

\item

  For the $\gamma < 0$ solutions, $\rho \to 0$ in the IR while $\chi$ 
  approaches some fixed value, $\widetilde \chi$, different from zero. 
  From equation \reef{ooohlovely}, this translates into some specific 
  value of $v > v_e$, which we shall denote as $w$. This is a 
  particular point (or ring of points) on the moduli space, and 
  although supergravity diverges there, we have the right to ask 
  physical questions. 
   
  This is where the probe physics comes in: {\it The metric that the 
    probe sees is perfectly smooth at $w$}.  We can evaluate the gauge 
  theory coupling as seen by the probe there using \reef{logarithm} 
  even though supergravity is singular. Deep enough in the IR, the 
  result is proportional to $N\log(w/ v_e)$, just as it should be for 
  the low energy effective gauge theory, at position/vev $w$ on the moduli 
  space.

\item For the $\gamma = 0$ solutions, $\chi$ diverges, and this is at 
  $v=v_e$. Supergravity has diverged and we appeal again to the 
  physics of the probe to help us. We see that the tension of the 
  probe drops to zero there: the gauge coupling diverges 
  logarithmically as seen in equation \reef{logarithm}. This is the 
  enhan\c con, and the radius $v=v_e$ is the locus of points where 
  there are new massless degrees of freedom in the gauge theory.

\end{itemize} 
 
We should digress here. The latter fact is borne out by the
realisation that the mass of a particle made by  a
macroscopic string stretched orthogonally to the D3--branes along the
$r$ direction (which make dyonic bound states in the gauge theory) has
a chance to drop to zero at $v=v_e$, since (by an analogous
computation to that done for the D3--brane) it is proportional to
\begin{equation} 
{\cal E}=\int_r^{r_e}dr\, F(r)(G_{rr}G_{tt})^{1/2}= 
\int_r^{r_e} dr\, F(r){\cosh^{1/2}(2\chi)\over\sinh(2\chi)}k\rho^4\ . 
\labell{tense} 
\end{equation} 
The part of the integrand not involving $F(r)$ is going to zero, as
its square is proportional to the D3--brane probe's $G_{vv}$. The
prefactor function $F(r)$ depends upon the type of macroscopic string
under investigation. For the D--string,
$F_{0,1}(r)=(2\pi\alpha^\prime)^{-1}e^{-\Phi/2}$, but it can be more
complicated for other strings, since their tension is controlled
by\cite{john}:
\begin{equation} 
F_{p,q}(r)={1\over 2\pi\alpha^\prime}\left[e^\Phi(qC_{(0)}-p)^2+ 
q^2e^{-\Phi}\right]^{1/2}\ . 
\end{equation} 
Inserting this into equation \reef{tense} essentially yields a  
formula for the mass of a $(p,q)$ dyon in the gauge theory as a 
function of $v$, as it is made of a bound state of $p$ fundamental 
strings and $q$ D1--branes ending on the D3--branes, whose tension is 
given by \reef{tense}.  These dyons will be mapped into one another by 
the $SL(2,\IZ)$ duality group bequeathed to the low energy theory by 
the parent type~IIB string theory. 
 
Given that we are in a non--trivial background, there is the 
possibility that divergent enough $F(r)$ might appear, to save some 
bound states from going massless, and it is  interesting to 
compute just which of the $(p,q)$ do, generalising the structures in 
ref.\cite{seibwitt}. In fact, (for the moduli space choice 
$\theta{=}\pi/2$), equations \reef{axidill} and \reef{relations} yield: 
\begin{eqnarray} 
&&e^{-\Phi}={2\cosh(2\chi)\over\cosh^2(2\chi)(1 
+\cos 2\phi)+1-\cos 2\phi} \ ,\,\,{\rm and}\nonumber\\ 
&&C_{(0)}={\sinh^2(2\chi)\sin 2\phi\over\cosh^2(2\chi)(1 
+\cos 2\phi)+1-\cos 2\phi}\ . 
\labell{axidilaton} 
\end{eqnarray} 
Furthermore, taking the limit appropriate to $\gamma=0$, ({\it i.e.,}
$\chi\to\infty$ and $\rho\to 0$), we see that because of the large
power of $\rho$ present in equation \reef{tense}, {\it all} of the
types of strings/dyons are tensionless on the enhan\c con, since
nothing can save ${\cal E}(r)$ from going to zero as $r\to
r_e$. This is generalisation of the features in ref.\cite{seibwitt}
which is consistent with the fact that constituent D3--branes of the
supergravity solution have spread out to form this ring, and so all
strings stretched between our probe and the background D3--branes will
give zero mass states in the limit.

\begin{figure}[ht]  
\centerline{\psfig{figure=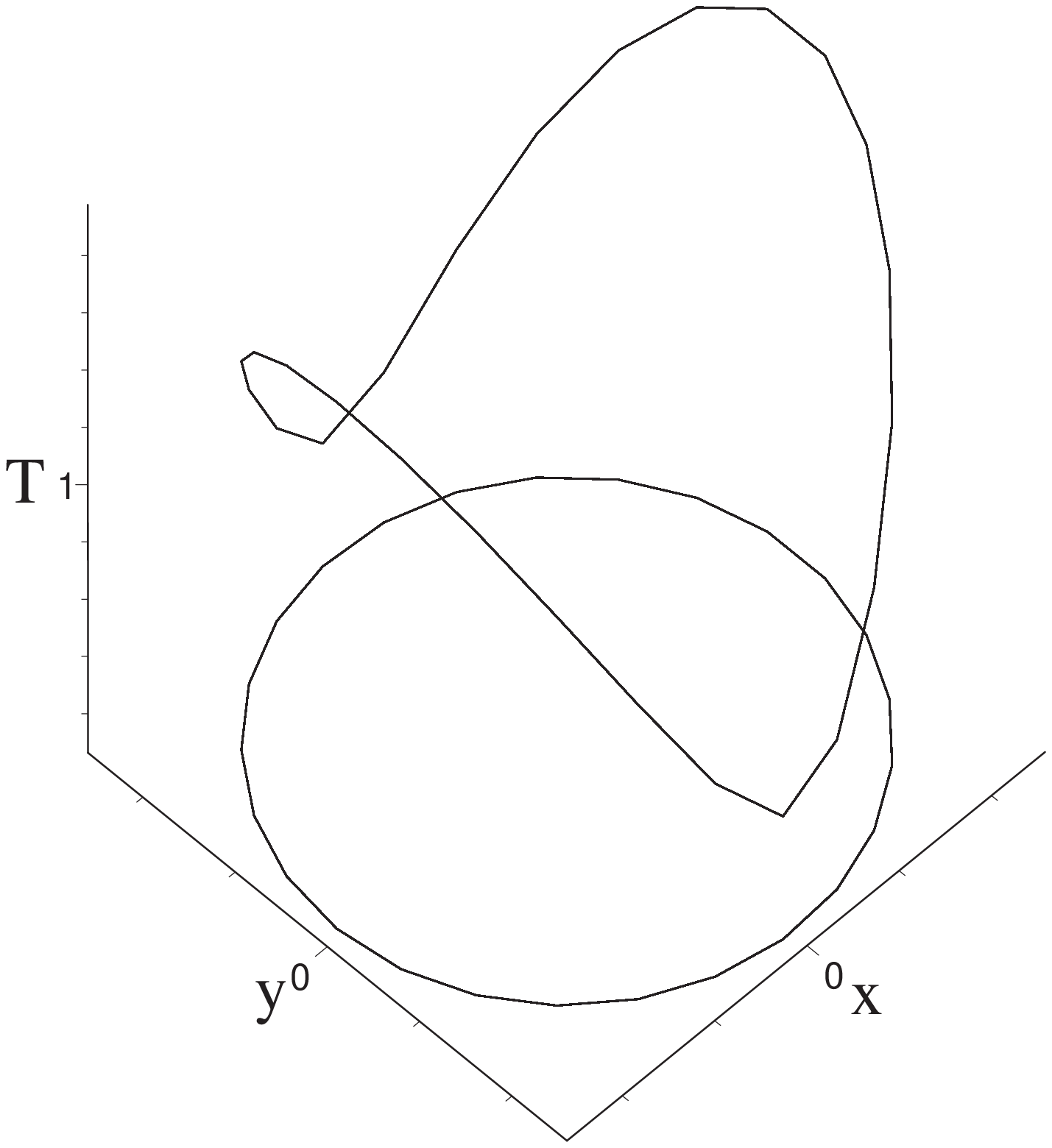,width=2.0in,height=1.8in}
\psfig{figure=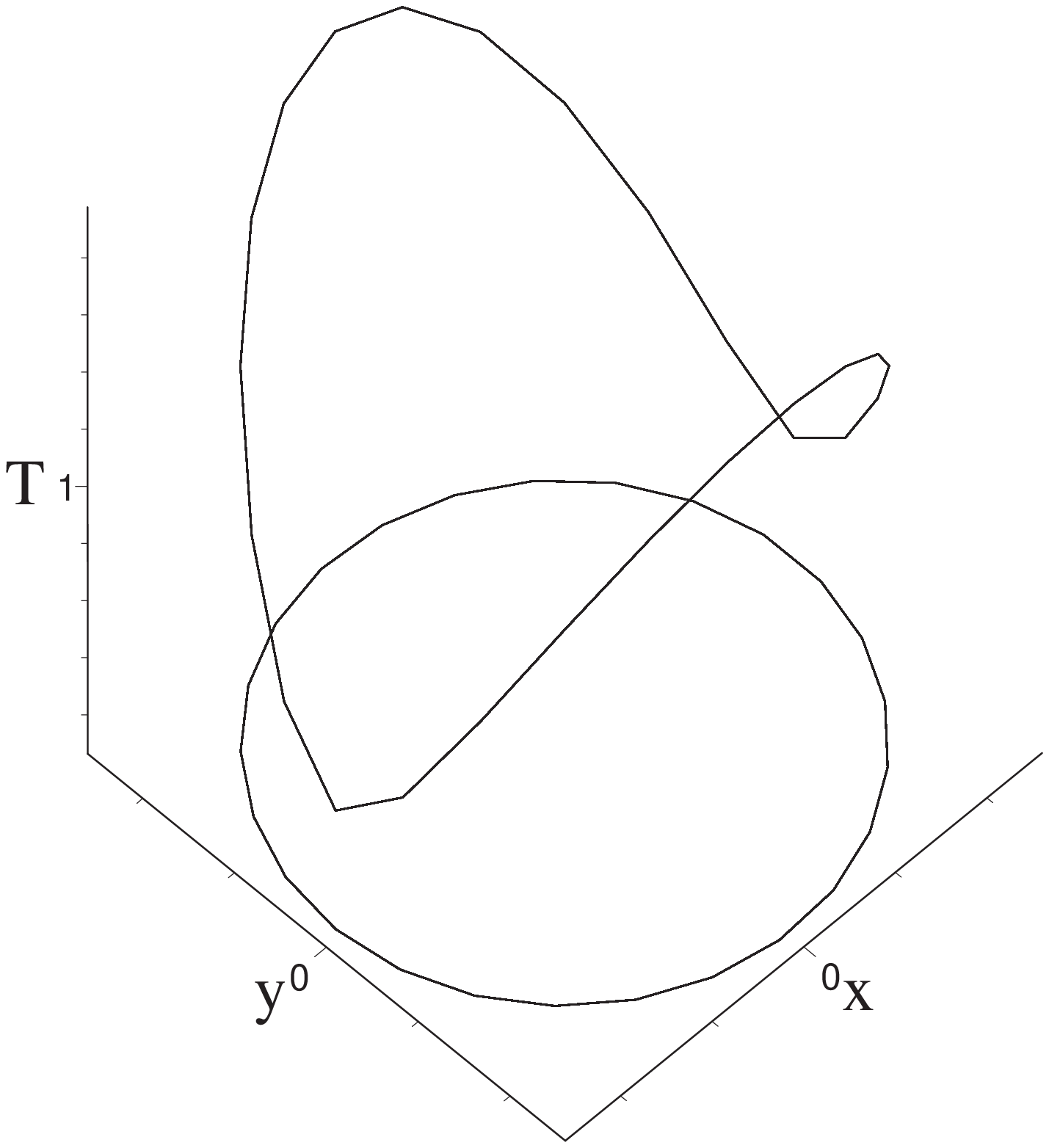,width=2.0in,height=1.8in}
\psfig{figure=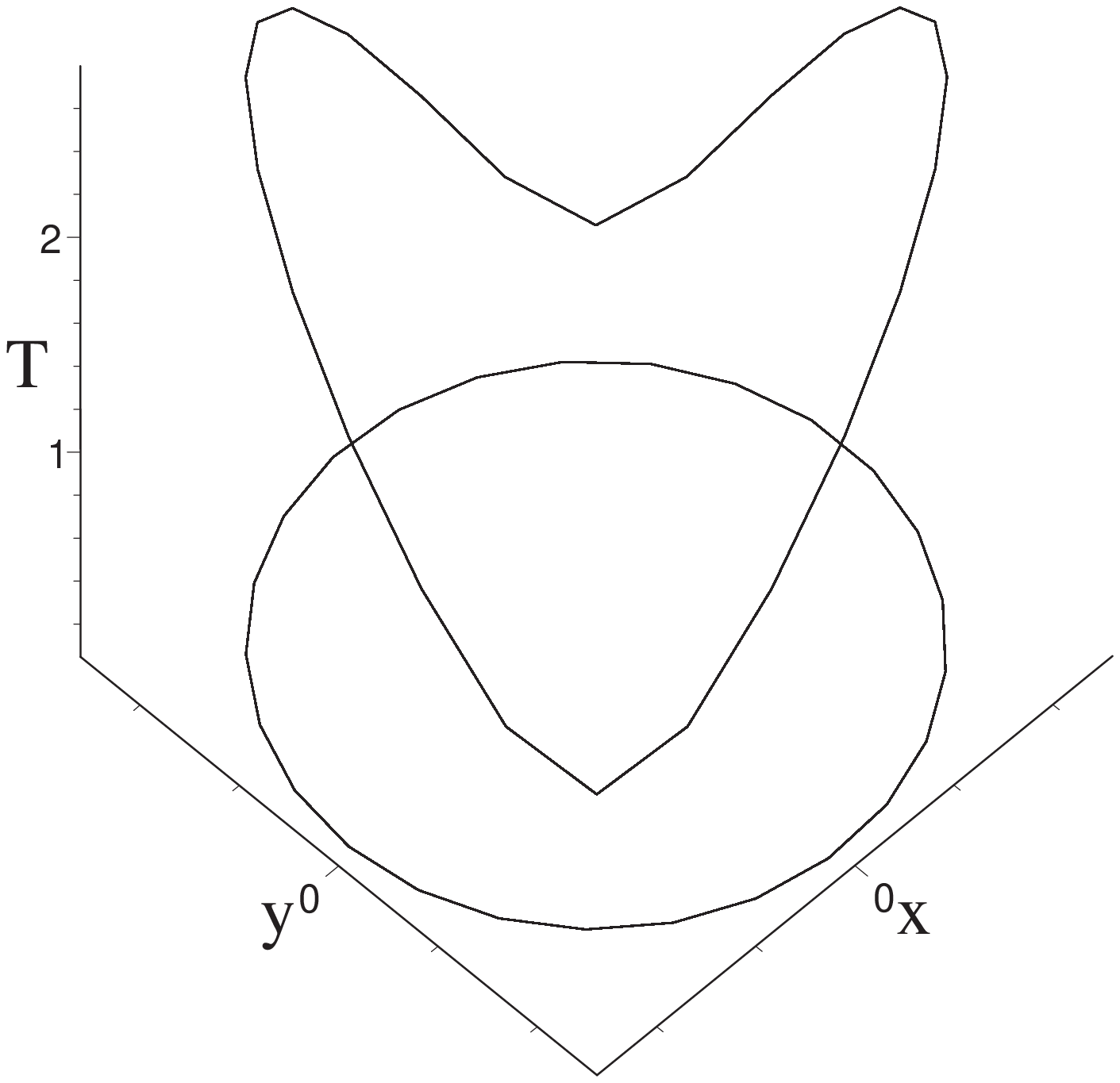,width=2.0in,height=1.7in}} 
\caption{\small The functional dependence of the tension of $(1,0)$,
  $(0,1)$, and $(1,1)$ strings as one goes around a generic circle. At
  the  enhan\c con, any point on the circle will have a lightest
  string/dyonic state.}
\label{wavy}  
\end{figure}

A tantalising feature of the special family of vacua represented by
the enhan\c con is that the dilaton and R--R scalar change as one goes
around the circle in $\phi$, moving along the enhan\c con ring: the
$(p,q)$--``flavour'' of the vacuum changes! Since the overall factor
not involving $F_{p,q}(r)$ is common to all, it seems intuitively
clear that there is a ``lightest'' dyon state among those which go
massless, which characterises the vacuum at a given value of~$\phi$.
This is easy to see by plotting some samples of the behaviour of
$F_{p,q}(\phi)$ for a value of $v$ away from the enhan\c con, as we
have done in figure~\ref{wavy}.

Furthermore, it is clear that the duality which descends from the
$SL(2,\IZ)$ of the type~IIB string plays a beautiful role here, by
exchanging the different vacua at different points on the circle.  For
example, the points $\phi=\{0,\pi/2,\pi,3\pi/2\}$, forming the
principal points on the compass, are special in that $C_{(0)}=0$.  So
a $(p,q)$ dyon's mass is controlled by the familiar Pythagorean form:
$\sqrt{p^2e^\Phi+q^2 e^{-\Phi}}$. However, as we go around the four
points, the dilaton toggles between $\Phi$ and $-\Phi$. These are
clearly vacua which are related by the action of the $S$ generator of
the $SL(2,\IZ)$. We should now return to our discussion of the flows.
 
\begin{itemize} 
\item 
   
  The $\gamma > 0$ curves appear to give behaviour which is 
  irretrievably unphysical from both the point of view of the 
  supergravity and the probe. This conclusion is in line with the 
  previous suggestions of refs.\cite{pw1,gub2,bs2}, but we make a 
  more detailed suggestion: In it natural to suppose that these flows 
  attempt to take us {\it behind} the enhan\c con ring to get to the 
  limiting IR physics, where the tension of the probe would need to be 
  (unphysically) negative in order to preserve supersymmetry. We 
  propose, in the spirit of ref.\cite{jpp}, that these are not 
  physical solutions for that region.  Excising that part of the 
  supergravity solution and (perhaps) replacing it with flat space 
  seems prudent.  Whether or not this gives a description of a part of 
  gauge theory moduli space possibly hidden inside the enhan\c con is 
  a matter for future work. 
 
\end{itemize}

We note again that we also have a full strong/weak coupling duality
group acting on the low energy gauge theory. (It is inherited from the
$SL(2,\IZ)$ symmetry of the D3--brane within type~IIB string theory,
as discussed in section \ref{gaugetheory}). This means that although
the gauge coupling diverges on the probe at the enhan\c con, there is
a dual description to a low energy theory with coupling $1/\tau$ and
complex modulus field $V_D$. This duality may be interesting to study,
although the dual vacua reached thus will also have a full tower of
massless dyons, and may require altogether new techniques to study
them, since some non--locality certainly arises.  Nevertheless, this
might be a promising new handle on the issue\cite{me} of finding an
effective low--energy description of the physics near the enhan\c con,
and may also shed light on whether one can make sense of the stringy
physics behind it in terms of gauge theory.

\bigskip 
 
\section*{Added Note}  
We have been informed\cite{joecomment} that Alex Buchel, Amanda Peet, 
and Joe Polchinski have done independent work on the issue of probing 
the solution of ref.\cite{pw1}, and understanding the role of the 
enhan\c con in this context\cite{bpp}.

\section*{Acknowledgements} 
We thank Juan Maldacena, Rob Myers, Joe Polchinksi, Nick Warner and 
Alberto Zaffaroni for conversations and comments.  N.E. is grateful to 
PPARC for the sponsorship of an Advanced Fellowship. C.V.J. would like 
to thank the Aspen Centre for Physics for hospitality while this work 
was completed. M.P. is partially supported by INFN and MURST, and by 
the European Commission TMR program ERBFMRX-CT96-0045, wherein she is 
associated to Imperial College, London, and by the PPARC SPG grant 
613. This paper is report number DTP/00/69 at the CPT, Durham. 
 


\end{document}